\documentclass[prl,aps,showpacs,twocolumn,floatfix]{revtex4}
\usepackage{epsfig} \usepackage{graphics} \usepackage{bm}
\usepackage{amssymb}
\usepackage{graphicx}
\begin{document}

\preprint{Lebed-Rapids-LN}

\title{Reentrant orbital effect against superconductivity in
the quasi-two-dimensional superconductor NbS$_2$}

\author{A.G. Lebed$^*$}

\affiliation{Department of Physics, University of Arizona, 1118 E.
4-th Street, Tucson, AZ 85721, USA}

\begin{abstract}
We derive integral equation for superconducting gap, which
takes into account the quantum nature of electron motion in a
parallel magnetic field in a quasi-two-dimensional (Q2D)
superconductor in the presence of a non-zero perpendicular field
component. By comparison of our theoretical results with the recent
experimental data obtained on the NbS$_2$, we show that the
orbital effect against superconductivity  partially destroys
superconductivity in the so-called Ginzburg-Landau area of this
Q2D conductor, as expected. Nevertheless, at relatively
high magnetic fields, $H \simeq 15 \ T$, the orbital effect
starts to improve the Fulde-Ferrell-Larkin-Ovchinnikov phase in the
NbS$_2$, due to the quantum nature of electron motion in a
parallel magnetic field. In our opinion, this is the most clear
demonstration that the orbital effect against superconductivity
in a parallel magnetic field has a reversible nature.

\end{abstract}

\pacs{???, 74.25.Op, 74.25.Ha}

\maketitle

It is well known that superconductivity at zero temperature is
usually destroyed in any superconductor by either the upper
orbital critical magnetic field, $H_{c2}(0)$, or the so-called
Clogston paramagnetic limiting field, $H_p$ [1]. These are due to
the fact that, in the traditional singlet Cooper pair, the
electrons possess opposite momenta and opposite spins. By present
moment, there are also known several superconducting phases, which
can exist above $H_{c2}(0)$ and $H_p$. Indeed, the paramagnetic
limit, $H_p$, can be absent for some triplet superconductors (see,
for example, UBe$_2$ [2-5]). Alternatively, for singlet
superconductivity, the superconducting phase can exceed the
Clogston limit by creating the non-homogeneous
Fulde-Ferrell-Larkin-Ovchinnikov (FFLO or LOFF phase) [6,7]. On
the other hand, if $H_{c2}(0)$ tries to destroy superconductivity,
then quantum effects of electron motion in a magnetic field can,
in principle, restore it as the Reentrant Superconducting (RS)
phase [8-17]. Although there are numerous experimental results
[18-31], confirming the existence of the FFLO phase in several Q2D
superconductors, there exist only a few experimental works [2-5],
where the presumably RS phase revives in ultrahigh magnetic fields due to
quantum effects of electron motion in a magnetic field in one
compound - UBe$_2$. On the other hand, the above mentioned unique RS
phenomenon has been theoretically predicted for a variety of Fermi
surfaces: for Q1D [8-10], for isotropic 3D [11], and for Q2D
superconductors [12-17].

Recently, the FFLO phase has been found by Lortz and collaborators
in the Q2D compound NbS$_2$ in a parallel magnetic filed [31]. The
peculiarity of this work is that at relatively low magnetic fields
(i.e., in the Ginzburg-Landau (GL) area [1]) the orbital effect
of the field partially destroys superconductivity but, at high
magnetic fields, everything looks like there is no any orbital
effect against superconductivity.
The aim of our Letter is to show that these happen due to the
reentrant nature of the quantum effects of electron motion in a
parallel magnetic field, theoretically predicted and considered in Refs.
[12-17]. To this end, we derive the so-called gap equation,
determining the upper critical field in slightly inclined magnetic
field, which directly takes into account quantum effects of
electron motion in a parallel magnetic field. The physical origin
of the above mentioned quantum effects [8,12] is related to the
Brag reflections from the Brillouin zone boundaries during
electron motion in a parallel magnetic field. To compare
the obtained results with the existing experimental
data, we derive the gap equation both for a strictly parallel
magnetic field and for a magnetic field with some perpendicular
component. The latter is derived, for the best of our knowledge,
for the first time. We use comparison of these equations with
experimental data [31] to extract the so-called GL
coherence lengths and in-plane Fermi velocity. These allow us to
show that, indeed, in the magnetic fields range $H \simeq 15 \
T$, quantum effects are very strong and completely suppress the
orbital effect against superconductivity. As a result, the FFLO
phase appears with the transition temperature value like for a pure
2D superconductor, which satisfies the experimental situation in
NbS$_2$ [31]. In our opinion, this is the first firm demonstration
of a reversible nature of the orbital effect against
superconductivity [8-17].

First, let us explain what physical picture corresponds to phase
diagram in a parallel magnetic field in the Q2D conductor
NbS$_2$, using qualitative arguments. Below, we consider a Q2D
conductor with the following electron spectrum, which is an
isotropic one within the conducting plane:
\begin{equation}
\epsilon({\bf p})= \frac{ (p^2_x + p^2_y)}{2m} - 2 t_{\perp}
\cos(p_z d), \ \ t_{\perp} \ll \epsilon_F = \frac{p^2_F}{2m},
\end{equation}
where $m$ is the in-plane electron mass, $t_{\perp}$ is
the integral of the overlapping of electron wave functions in a
perpendicular to the conducting planes direction; $\epsilon_F$ and
$p_F$ are the Fermi energy and Fermi momentum, respectively;
$\hbar \equiv 1$. In a parallel to the conducting planes magnetic
field which is applied along ${\bf y}$ axis,
\begin{equation}
{\bf H} = (0,H_{\parallel},0) \ ,
\end{equation}
it is convenient to choose vector potential of the field in the
following form:
\begin{equation}
{\bf A} = (0,0,-H_{\parallel}x) \ .
\end{equation}
The quasi-classical electron motion along open orbits in the
parallel magnetic field (2) is due to the action of
z-component of the Lorentz force:
\begin{equation}
\frac{d p_z}{dt} = \frac{e v_F \cos (\phi) H_{\parallel}}{c}  \
, \ \ p_zd = \omega_{\parallel} \cos(\phi) t,
\end{equation}
where the typical cyclotron frequency of electron motion along open orbits
is
\begin{equation}
\omega_{\parallel} = \frac{e v_F d H_{\parallel}}{c},
\end{equation}
with the polar angle, $\phi$, being counted from  the ${\bf x}$
axis. For open electron orbits,
\begin{equation}
v_z(p_z)= 2t_{\perp} \frac{ d[\cos(p_z d)]}{dp_z} = -2 t_{\perp}d
\sin (p_zd),
\end{equation}
therefore, electron motion perpendicular to the
conducting layers in a real space can be represented as
\begin{eqnarray}
&z(\phi, t) = \frac{l_{\perp}} {\cos(\phi)} \
\cos[\omega_{\parallel}\cos(\phi) t],
\nonumber\\
&l_{\perp} = 2 t_{\perp} d / \omega_{\parallel}.
\end{eqnarray}

Let us discuss Eq.(7) for electron motion between the conducting
layers in the Q2D superconductor NbS$_2$. As we show below, in low
magnetic fields (i.e., in the GL region [1]), all electron
trajectories have magnitudes of motion (7) bigger than the
inter-plane distance, $l_{\perp} \geq d$. Therefore, in low
enough magnetic fields, the Meissner destructive currents against
superconductivity are effective and, thus,
superconductivity is partially destroyed by the orbital effects in
accordance with the GL theory [1]. On the other hand, as we also
show below, at magnetic fields, where the FFLO phase exists, $H
\simeq 15 \ T$, the significant amount of electrons possesses
the quasi-classical trajectories with magnitude of the order
or less than the interatomic distance, $l_{\perp} \ll d$.
In the latter case, the orbital effect against superconductivity becomes
small [8-17] in parallel magnetic fields [12-17] since electrons are
almost localized on the conducting layers. If there were no the Pauli
paramagnetic effects against superconductivity, then the RS phase
would appear in high magnetic fields. Nevertheless, in a reality
the Pauli spin-splitting paramagnetic effects lead to the appearance
of the LOFF phase, which is eventually destroyed by the field.

Let us consider slightly inclined with respect to the conducting
planes magnetic field,
\begin{equation}
{\bf H} = (0,H_{\parallel},H_{\perp}) \ ,
\end{equation}
since the experiments in Ref.[31] are done both for the parallel
(2) and the slightly inclined (8) fields. For our
calculations, it is convenient to choose the following gauge,
where the vector-potential of the magnetic field (8) depends only
on coordinate $x$:
\begin{equation}
{\bf A} = (0,H_{\perp}x,-H_{\parallel}x) \ .
\end{equation}

To describe electron motion in the inclined magnetic field (8), we
make use of the so-called Peierls  substitution method [1]:
\begin{equation}
p_x = -i \frac{\partial}{\partial x}, \ \ p_y = -i
\frac{\partial}{\partial y} -\frac{e}{c}A_y, \ \ p_z= -i
\frac{\partial}{\partial z}-\frac{e}{c}A_z.
\end{equation}
As a result, the electron Hamiltonian in the magnetic field (8)
can be represented as:
\begin{eqnarray}
&&\hat H = \frac{1}{2m} \biggl[ -\biggl( \frac{\partial }{\partial x}
\biggl)^2 + \biggl(-i \frac{\partial}{\partial y} -\frac{e}{c} H_{\perp}
x\biggl)^2 \biggl] \nonumber\\ &&-2 t_{\perp} \cos \biggl(-i d
\frac{\partial}{\partial z} + \frac{edH_{\parallel}
x}{c} \biggl).
\end{eqnarray}
Let us introduce electron wave functions,
\begin{eqnarray}
&\Psi^{\pm}_{\epsilon, p_y, p_z}({\bf r}) = \exp[\pm i p^0_x(p_y) x]
\ \exp( i p_y y)
\nonumber\\
&\times \exp( i p_z z) \ \Psi^{\pm}_{\epsilon}(x;p_y,p_z) \ ,
\end{eqnarray}
where the two-component electron momentum vector, $[p^0_x(p_y),p_y]$, is
located on the
$2D$ Fermi surface (FS),
\begin{equation}
[p^0_x(p_y)]^2 + p^2_y = p_F^2,
\end{equation}
and where (+) and (-) stands for $[p^0_x(p_y) >0]$ and
$[p^0_x(p_y)<0]$, respectively. It is important that, for the main
part of the $2D$ FS, the following conditions of the quasiclassical
motion in the magnetic field are valid:
\begin{equation}
p^0_x(p_y), \ p_y \sim p_F \gg e H_{\perp} / p_F c.
\end{equation}
It is possible to prove that, in this case, we can rewrite the
Schr\"{o}dinger equation (11) for the wave functions
$\Psi^{\pm}_{\epsilon}(x;p_y,p_z)$ in Eq.(12) as
\begin{eqnarray}
&\biggl[\mp i v_F \cos (\phi) \frac{d}{dx} -\omega_{\perp} p_y x  -
2t_{\perp} \cos\biggl(p_z d + \frac{\omega_{\parallel} \
x}{v_F}\biggl) \biggl]
\nonumber\\
&\times \Psi^{\pm}_{\epsilon}(x; p_y, p_z)= \delta \epsilon \
\Psi^{\pm}_{\epsilon}(x; p_y, p_z),
\end{eqnarray}
where:
\begin{equation}
\omega_{\perp}= \frac{eH_{\perp}}{mc}, \ \ \delta \epsilon =
\epsilon - \epsilon_F.
\end{equation}
It is important that Eq.(15) can be solved exactly. As a result,
we obtain:
\begin{eqnarray}
&&\Psi^{\pm}_{\epsilon}(x;p_y,p_z)= \exp \biggl[\pm i\frac{\delta
\epsilon \ x}{v_F \cos (\phi)}\biggl] \exp \biggl[\pm i
\frac{\omega_{\perp} p_F \sin (\phi) x^2}{2 v_F \cos (\phi)}\biggl]
\nonumber\\
&&\times \exp \biggl[\pm i \frac{2t_{\perp}}{\omega_{\parallel}
\cos (\phi)} \sin \biggl( p_z d +\frac{\omega_{\parallel}
x}{v_F} \biggl) \biggl].
\end{eqnarray}

 At this point, we define electron Matsubara Green's functions in the
 mixed, $(x;p_y,p_z)$, representation using Refs. [32,33],
\begin{equation}
(i \omega_n - \hat H) g_{i \omega_n}^{\pm}(x,x_1;p_y,p_z)=
\delta(x-x_1),
\end{equation}
and obtain:
\begin{eqnarray}
&g^{\pm}_{i \omega_n}(x,x_1;p_y,p_z)= -i \frac{sgn \ \omega_n}{v_F
\cos \phi} \exp \biggl[\mp \frac{\omega_n (x-x_1)}{v_F \cos \phi}
\biggl]
\nonumber\\
&\times \exp \biggl\{\pm i \frac{2 t_{\perp}}{\omega_{\parallel} \cos \phi}
\biggl[\sin \biggl(p_z d +\frac{\omega_{\parallel} \
x}{v_F}\biggl) - \sin \biggl(p_z d +\frac{\omega_{\parallel}
x_1}{v_F}\biggl) \biggl] \biggl\}
\nonumber\\
&\times \exp \biggl[\mp i \frac{\omega_c p_F \sin \phi (x^2-x^2_1)}{2 v_F
\cos \phi}\biggl], \ \ \pm \omega_n (x-x_1)>0.
\end{eqnarray}

Let us derive the so-called gap equation for superconducting order
parameter, $\Delta(x)$, which defines the upper critical magnetic field,
destroying superconductivity. To this end, we use the linearized
Gor'kov's equations for a non-uniform superconductivity [32,34],
\begin{eqnarray}
&&\Delta(x)= U \ T \sum_{|\omega_n|< \Omega} \int dp_y \int dp_z
\int dx_1 \ \Delta(x_1)
\nonumber\\
&&\times g^+_{i\omega_n}(x,x_1;p_y,p_z) \
g^-_{-i\omega_n}(x,x_1;-p_y,-p_z),
\end{eqnarray}
where the constant $U$ corresponds to the electron-electron
interactions. Note that below we consider the case of
$s$-superconductivity.
 As a result of straightforward but lengthly calculations, we obtain
\begin{eqnarray}
&\Delta(x) = U \int^{\pi}_{-\pi} \frac{d \phi}{2 \pi}
\int_{|x-x_1|>\frac{v_F |\cos \phi|}{\Omega}} \frac{2\pi T
dx_1}{v_F |\cos \phi| \sinh \biggl [\frac{2 \pi T |x-x_1|}{v_F
|\cos \phi|}\biggl]}
\nonumber\\
&\times J_0\biggl\{ \frac{8 t_{\perp}}{\omega_{\parallel}|\cos \phi|} \
\sin \biggl[\frac{\omega_{\parallel} (x-x_1)}{2 v_F} \bigg]
\sin \biggl[\frac{\omega_{\parallel} (x+x_1)}{2 v_F} \bigg]\biggl\}
\nonumber\\
&\times\cos \biggl[
\frac{\omega_{\perp} p_F \sin \phi (x^2-x^2_1)}{v_F \cos \phi}\biggl]
\ \Delta(x_1),
\end{eqnarray}
where $\Omega$ is a cut-off energy.
Note that gap Eq.(21) is rather general. First of all, it takes into
account the quantum nature of electron motion in perpendicular to the
conducting planes direction [8,12], as discussed early. It is a
periodic one and, thus, results in a spatial periodicity of the Bessel
function, $J_0(...)$, in Eq.(21). Second, it takes a possibility of
in-plane electron motion in the magnetic field (8)
at the quasi-classical level, since we have linearized the Hamiltonian
with respect to $\omega_{\perp}$. What is not taken into account in Eq.(21)
is the Pauli paramagnetic spin-splitting effects, which can be added in a
trivial way [12].
As a result,
\begin{eqnarray}
&\Delta(x) = U \int^{\pi}_{-\pi} \frac{ d \phi}{2 \pi}
\int_{|x-x_1|> \tilde d|\cos \phi|} \frac{2\pi T
dx_1}{v_F |\cos \phi| \sinh \biggl [\frac{2 \pi T |x-x_1|}{v_F
|\cos \phi|}\biggl]}
\nonumber\\
&\times J_0\biggl\{ \frac{8 t_{\perp}}{\omega_{\parallel}|\cos \phi|} \
\sin \biggl[\frac{\omega_{\parallel} (x-x_1)}{2 v_F} \bigg]
\sin \biggl[\frac{\omega_{\parallel} (x+x_1)}{2 v_F} \bigg]\biggl\}
\nonumber\\
&\times\cos \biggl[
\frac{\omega_{\perp} p_F \sin \phi (x^2-x^2_1)}{v_F \cos \phi}\biggl]
\cos \biggl[ \frac{2 \mu_B H (x-x_1)}{v_F \cos \phi} \biggl] \ \Delta(x_1),
\end{eqnarray}
where $\mu_B$ is the Bohr magneton, $\tilde d=v_F/\Omega$,
$H=\sqrt{H^2_{\parallel}+H^2_{\perp}}$.

Below, we will use Eq.(22) in two case: for a parallel magnetic field
both in the GL and RS regions,
\begin{eqnarray}
&\Delta(x) = U \int_{-\pi}^{\pi} \frac{d \phi}{2 \pi}
\int_{|x-x_1|> \tilde d |\cos \phi|} \frac{2\pi T
dx_1}{v_F |\cos \phi| \sinh \biggl [\frac{2 \pi T |x-x_1|}{v_F
|\cos \phi|}\biggl]}
\nonumber\\
&\times J_0\biggl\{ \frac{8 t_{\perp}}{\omega_{\parallel}|\cos \phi|} \
\sin \biggl[\frac{\omega_{\parallel} (x-x_1)}{2 v_F} \bigg]
\sin \biggl[\frac{\omega_{\parallel} (x+x_1)}{2 v_F} \bigg]\biggl\}
\nonumber\\
&\times
\cos \biggl[ \frac{2 \mu_B H_{\parallel} (x-x_1)}{v_F \cos \phi} \biggl] \ \Delta(x_1),
\end{eqnarray}
and for the presence of both parallel and perpendicular components of the
magnetic field (8) in the GL region, where quantum effects of electron
motions are not essential,
\begin{eqnarray}
&\Delta(x) = U \int_{-\pi}^{\pi} \frac{d \phi}{2 \pi}
\int_{|x-x_1|> \tilde d |\cos \phi|} \frac{2\pi T
dx_1}{v_F |\cos \phi| \sinh \biggl [\frac{2 \pi T |x-x_1|}{v_F
|\cos \phi|}\biggl]}
\nonumber\\
&\times J_0 \biggl[ \frac{2 t_{\perp} \omega_{\parallel}
(x^2-x^2_1)}{v^2_F|\cos \phi|} \biggl]
\nonumber\\
&\times \cos \biggl[
\frac{\omega_{\perp} p_F \sin \phi (x^2-x^2_1)}{v_F \cos \phi}\biggl]
\cos \biggl[ \frac{2 \mu_B H(x-x_1)}{v_F \cos \phi} \biggl] \ \Delta(x_1).
\end{eqnarray}
[As we mentioned above, Eq.(24) does not take into account the quantum
effects of electron motion in a parallel magnetic field, therefore, $\sin(...)$
functions in Eq.(23) are replaced by their arguments in the Bessel
function in Eq.(24) (see Refs. [36-38].)

Let us calculate the low-field GL slope of a parallel magnetic field, where
the quantum effects of electron motion as well as the Pauli paramagnetic
spin-splitting
effects are small. To this end, as a starting point, we have to consider the
following integral equation,
\begin{eqnarray}
&\Delta(x) = U \int_{-\pi}^{\pi} \frac{d \phi}{2 \pi}
\int_{|x-x_1|> \tilde d|\cos \phi|} \frac{2\pi T
dx_1}{v_F |\cos \phi| \sinh \biggl [\frac{2 \pi T |x-x_1|}{v_F
|\cos \phi|}\biggl]}
\nonumber\\
&\times J_0 \biggl[ \frac{2 t_{\perp} \omega_{\parallel}
(x^2-x^2_1)}{v^2_F|\cos \phi|} \biggl] \ \Delta(x_1),
\end{eqnarray}
and to introduce the more convenient variables:
\begin{equation}
x_1-x= z \cos \phi , \ \ x_1 = x +z \cos \phi , \ \ x_1+x=2x+z
\cos \phi.
\end{equation}
After simple transformations, integral Eq.(25) becomes a "non-divergent" one
and can be rewritten as:
\begin{eqnarray}
&\Delta(x) = g \int_{-\pi}^{\pi} \frac{d \phi}{2 \pi} \int^{\infty}_{\tilde d}
\frac{2 \pi T d z}{v_F
\sinh \biggl( \frac{ 2 \pi T z}{ v_F} \biggl) }
\nonumber\\
&\times J_0 \biggl\{ \frac{2 t_{\perp} \omega_{\parallel}}{v^2_F } [z(2x+z
\cos \phi)]\biggl\}  \ \Delta(x +z \cos \phi),
\end{eqnarray}
where $g$ is the electron coupling constant.
Here, we derive the GL slope of the parallel upper critical
magnetic field from Eq.(27). To this end, we expend the Bessel
function and the superconducting gap with respect to small parameter,
$z \ll v_F/(\pi T_c)$:
\begin{eqnarray}
&&J_0 \biggl\{ \frac{2 t_{\perp} \omega_{\parallel}}{v^2_F} [z (2x + z \cos
\phi)] \biggl\} \approx 1-\frac{4 t^2_{\perp}
\omega^2_{\parallel}}{v^4_F} x^2 z^2 \ ,
\nonumber\\
&&\Delta(x+z \cos \phi) \approx \Delta(x)+ \frac{1}{2} z^2 \cos^2 (\phi) \
\frac{d^2 \Delta(x)}{dx^2}.
\end{eqnarray}
The next step is to substitute Eq. (28) into integral (27)
and to average over angle $\phi$. As a result, we obtain:
\begin{eqnarray}
&&\Delta(x) \biggl[ \frac{1}{g} - \int_{\tilde d}^{\infty}  \frac{2 \pi T d
z}{v_F \sinh \biggl( \frac{ 2 \pi T z}{ v_F} \biggl)} \biggl]
\nonumber\\
&&-\frac{1}{4} \frac{d^2 \Delta(x) }{dx^2} \int_0^{\infty} \frac{2
\pi T_c z^2 d z}{v_F \sinh \biggl( \frac{ 2 \pi T_c z}{ v_F}
\biggl)}
\nonumber\\
&&+ x^2 \Delta(x) \frac{4t^2_{\perp}\omega^2_c}{v^4_F}
\int_0^{\infty} \frac{2 \pi T_c z^2 d z}{v_F \sinh \biggl( \frac{
2 \pi T_c z}{ v_F} \biggl)} = 0 \ ,
\end{eqnarray}
where $T_c$ is superconducting transition temperature in the
absence of a magnetic field, which satisfies the equation:
\begin{equation}
 \frac{1}{g} = \int_{\tilde d}^{\infty}  \frac{2 \pi T_c d z}{v_F
\sinh \biggl( \frac{ 2 \pi T_c z}{ v_F} \biggl)} \ .
\end{equation}
Here, we also take into account that [35]:
\begin{equation}
\int^{\infty}_0 \frac{x^2 dx}{\sinh(x)}  = \frac{7}{3} \zeta(3) \ ,
\end{equation}
where $\zeta(x)$ is the Riemann zeta-function, and introduce
the parallel and perpendicular GL coherence lengths,
\begin{equation}
\xi_{\parallel} = \frac{\sqrt{7 \zeta(3)}v_F}{4 \sqrt{2} \pi T_c}, \ \ \
\xi_{\perp} = \frac{\sqrt{7 \zeta(3)} t_{\perp} c^*}{2 \sqrt{2} \pi T_c},
\end{equation}
correspondingly.
Now differential gap Eq.(29) can be rewritten as:
\begin{equation}
- \xi^2_{\parallel} \frac{d^2 \Delta(x)}{dx^2}
+ \biggl(\frac{2\pi H_{\parallel}}{\phi_0}\biggl)^2 \xi^2_{\perp} x^2 \Delta(x)
-\tau \Delta(x) = 0,
\end{equation}
where $\phi_0 = \frac{\pi c}{e}$ is the magnetic flux quantum,
$\tau=\frac{T_c-T}{T_c}$. It is important that the GL Eq.(33) can be
analytically solved [1] and expression for the GL upper critical magnetic
field slope can be analytically written:
\begin{equation}
H^{\parallel}_{c2} = \tau \biggl( \frac{\phi_0}{2 \pi
\xi_{\parallel} \xi_{\perp}} \biggl) = \tau \biggl[ \frac{8 \pi^2
c T^2_c}{7 \zeta(3) e v_F t_{\perp} c^*} \biggl] .
\end{equation}

Let us calculate the GL upper critical magnetic field for the slightly
inclined field (8). In this case, we can write Eq.(24) in new
variables (26) in the following way:
\begin{eqnarray}
&\Delta(x) = g \int_{-\pi}^{\pi} \frac{d \phi}{2 \pi} \int^{\infty}_{\tilde d}
\frac{2 \pi T d z}{v_F
\sinh \biggl( \frac{ 2 \pi T z}{ v_F} \biggl) }
\nonumber\\
&\times \cos \biggl[
\frac{e H_{\perp}}{c} \sin (\phi) \ z \ (2x + z \cos \phi)
 \biggl]
\nonumber\\
&\times J_0 \biggl\{ \frac{2 t_{\perp} \omega_{\parallel}}{v^2_F } [z(2x+z
\cos \phi)]\biggl\}  \ \Delta(x +z \cos \phi).
\end{eqnarray}
Using the same procedure as before and expending of the addition $\cos(...)$
function in Eq.(35) with respect to small magnetic field $H_{\perp}$, we
can obtain:
\begin{equation}
\frac{1}{H^2_{c2}(\alpha)}= \frac{\sin^2 (\alpha)}{(H^{\perp}_{c2})^2}+
\frac{\cos^2 (\alpha)}{(H^{\parallel}_{c2})^2},
\end{equation}
where
\begin{equation}
\tan \alpha = \frac{H_{\perp}}{H_{\parallel}}, \ \
H^{\perp}_{c2} = \tau \biggl( \frac{\phi_0}{2 \pi
\xi^2_{\parallel}} \biggl) = \tau \biggl[ \frac{16 \pi^2 c
T^2_c}{7 \zeta(3) e v_F^2} \biggl] .
\end{equation}

Note that the upper critical magnetic fields were measured in Ref.[31] for
$\alpha_0 =0$ and $\alpha_1 = 1^0$. Taking into account Eqs.(34),(36),(37)
and applying them to the experiment [31], we obtain:
\begin{eqnarray}
&H^{\parallel}_{c2}\simeq 32 \ T, \ H^{\perp}_{c2} \simeq 0.31 \ T , \
t_{\perp} \simeq 4 \ K,
\nonumber\\
&\xi_{\parallel}\simeq 320 \ \AA , \ \xi_{\perp}\simeq 3.1 \AA , \ v_F \simeq
1.5 \times 10^7 cm/s.
\end{eqnarray}
Moreover, using also Eq.(5) and the value $d \simeq 12 \ \AA$, we are able
to write that
\begin{equation}
\omega_c(H=1 \ T) \simeq 2 \ K.
\end{equation}
Here, we stress that in the GL area of NbS$_2$, where $H \leq 5 \ T$,
it follows from Eq.(7) that $l_{\perp}/d \geq 1$ and thus, the orbital effect
against superconductivity effectively decreases superconducting transition
temperature in agreement with the experiment [31]. On the other
hand, in the area of the existence of the FFLO phase (i.e., at $H \simeq
15 \ T$), $l_{\perp}/d \simeq 0.27$, which, as we show below, makes
the orbital destructive effect almost to disappear.

Let us now quantitatively consider the area of relatively high magnetic fields,
 $H \simeq
15 \ T$, where $l_{\perp}/d \ll 1$. In this case the orbital effect
against superconductivity can be considered as a perturbation in Eq.(23),
which can be written as:
\begin{eqnarray}
\frac{1}{g} = \int^{\pi}_{-\pi} \frac{d \phi}{2 \pi} \int^{\infty}_{\tilde d}
\frac{dz}{z}
\biggl\{ \biggl[1 - 2 \frac{(l_{\perp}/d)^2}{\cos^2 (\phi) } \sin^2
\biggl(\frac{\omega_{\parallel} z \cos \phi}{4v_F} \biggl) \biggl]
\nonumber\\
\times \cos \biggl( \frac{2 \mu_B H z}{v_F} \biggl)
\cos \biggl( \frac{2 \mu_B H z \cos \phi}{v_F} \biggl) \biggl\} .
\end{eqnarray}
Note that while deriving Eq.(40), we expand the Bessel functions in Eqs.(23)
and (27)
with respect to the small parameter, $l_{\perp}/d \ll 1$, consider the case
of small temperatures, $T=0$, and use the FFLO
solution for the superconducting gap [6]:
\begin{equation}
\Delta (x) = \Delta_0 \cos \biggl(\frac{2 \mu_B H z}{v_F} \biggl).
\end{equation}
It is important that the gap (41) is the solution of Eq.(40) for
$l_{\perp}$=0 and corresponds to the following FFLO critical magnetic
field [39]:
\begin{equation}
H_{FFLO}=\frac{\Delta_0}{\mu_B}= \frac{\pi T_c}{2 \gamma \mu_B},
\end{equation}
where $\gamma$ is the Euler's constant [35].
It is possible to show that, in our case, Eq.(40) results in the following
correction, $H^*_{FFLO}$, to the FFLO critical magnetic field (42):
\begin{eqnarray}
&&\frac{H_{FFLO}-H^*_{FFLO}}{H_{FFLO}}=  2 \frac{l^2_{\perp}}{d^2} \int^{\pi}_{-\pi}
\frac{d \phi}{2 \pi} \int^{\infty}_0
\frac{dz}{z}
\nonumber\\
&&\times \frac{\sin^2 [(\omega_{\parallel} z \cos \phi) /4]}{ \cos^2
(\phi)}
 \cos \biggl( \frac{2 \mu_B H z}{v_F} \biggl)
 \cos \biggl(\frac{2 \mu_B H z \cos \phi}{ v_F} \biggl) .
\end{eqnarray}

Numerical integration of Eq.(43) for $\omega_{\parallel}(H=1 \ T) = 2 \ K$ and
$2 \mu_B H (H = 1 T) = 1.35 \ K$
gives the following result:
\begin{equation}
\frac{H^*_{FFLO}-H_{FFLO}}{H_{FFLO}} = - 0.2 \ \frac{l^2_{\perp}}{d^2} .
\end{equation}
Finally, taking into account that, at $H \simeq 15 \ T$, $l_{\perp}/d \simeq
0.27$, we find that the relative change of the critical field of the appearance
of the FFLO phase is very small:
\begin{equation}
\frac{H^*_{FFLO}-H_{FFLO}}{H_{FFLO}} = - 0.015.
\end{equation}
It is important that this result is in agreement with the experimental data [31],
where the experimental curve for the FFLO phase is scaled with
experimental results obtained on the almost 2D superconductor
$\kappa$-(BEDT-TTF)$_2$Cu(NCS)$_2$, where the orbital effect
against superconductivity is extremely small at any magnetic fields
and the GL decrease
of superconductivity is not clear
experimentally observed [23]. [Note that the so-called Lawrence-Doniach
parameter, which defines a degree of "three-dimensionality" in a Q2D
superconductor [40,41], can be estimated as $\sqrt{2} \xi_{\perp}/d \leq 0.1$
for the $\kappa$-(BEDT-TTF)$_2$Cu(NCS)$_2$ and as
$\sqrt{2} \xi_{\perp}/d \simeq 0.4$ for the NbS$_2$, respectively. This means
that the Q2D
superconductor NbS$_2$ is much more "three-dimensional" than the
$\kappa$-(BEDT-TTF)$_2$Cu(NCS)$_2$ one.]

To summarize, we have demonstrated that the orbital effect against
superconductivity has a reversible nature in the Q2D conductor NbS$_2$
in a parallel magnetic field.
Indeed, as shown, at low enough magnetic fields, $H \leq 5 \ T$, the
orbital effect partially destroys superconductivity in an agreement with
the GL theory. On the other hand, as also demonstrated, at relatively
high magnetic fields, $H \simeq 15 \ T$, the orbital effect almost disappears
due to quantum nature of electron motion in a parallel magnetic field.

The author is thankful to N.N. Bagmet and Walter Lortz for useful
discussions.

$^*$Also at: L.D. Landau Institute for Theoretical Physics, RAS, 2
Kosygina Street, Moscow 117334, Russia.


\begin{references}

\bibitem{Abrikosov-1}
See, for example, book A.A. Abrikosov, {\it Fundamentals of Theory
of Metals} (Elsevier Science, Amsterdam, 1988).

\bibitem{2} S. Ran, C. Eckberg, Q.-P. Ding, Y. Furukawa, T. Metz, S.R. Saha,
I-Lin Liu, V.Zic, H. Kim, J. Paglione, and N.P. Butch {\it
Science} {\bf 365}, 684 (2019).

\bibitem{3}
S. Ran, I-Lin Liu, YunSuk Eo, D.J. Campbell, P.M. Neves, W.T.
Fuhrman, S.R. Saha, C. Eckberg, H. Kim, D. Graf, F. Balakirev, J.
Singleton, J. Paglione, and N.P. Butch {\it Nature Physics} {\bf
15}, 1250 (2019).


\bibitem{4} D. Aoki, A. Nakamura, F. Honda, DeXin Li, Y. Homma, Y. Shimizu,
Y.J. Sato, G. Knebel, J.-P. Brison, A. Pourret, D. Braithwaite, G.
Lapertot, Qun Niu, M. Valiska, H. Harima, and J. Flouquet, {\it J.
Phys. Soc. Jpn.} {\bf 88}, 043702 (2019).


\bibitem{5} G. Knebel, W. Knafo, A. Pourret, Qun Niu, M. Valiska, D. Braithwaite,
G. Lapertot, M. Nardone, A. Zitouni, S. Mishra, I. Sheikin, G.
Seyfarth,  J.-P. Brison, D. Aoki, and J. Flouquet, {\it J. Phys.
Soc. Jpn.} {\bf 88}, 063707 (2019).


\bibitem{25} A.I. Larkin and Yu.N. Ovchinnikov, {\it Sov. Phys. JETP} {\bf 20}, 762 (1965).


\bibitem{26} P. Fulde and R.A. Ferrell, {\it Phys. Rev.} {\bf 135}, A550
(1964).

\bibitem{8} A.G. Lebed, {\it JETP Lett.} {\bf 44}, 114 (1986).


\bibitem{9} N. Dupuis, G. Montambaux, and C.A.R. Sa de Melo,
{\it. Phys. Rev. Lett.} {\bf 70}, 2613 (1993).

\bibitem{10} A.G. Lebed and O. Sepper, {\it Phys. Rev.} {\bf B 90}, 024510 (2014).

\bibitem{Razolt} M. Razolt and Z. Tesanovic, {\it Rev. Mod. Phys} {\bf 64}, 709
(1992).



\bibitem{11} A.G. Lebed and K. Yamaji {\it Phys. Rev. Lett.} {\bf 80}, 2697 (1998).


\bibitem{12} A.G. Lebed, {\it J. of Supercond.} {\bf 12}, 453 (1999).


\bibitem{13} M. Miyazaki, K. Kishigi, and Y. Hasegawa {\it J. Phys. Soc. Jp.}
{\bf 67}, L2618, (1998).


\bibitem{14} V.P. Mineev {\it J. Phys. Soc. Jp.} {\bf 69}, 3371 (2000).


\bibitem{15} V.P. Mineev {\it JETP Lett.}, {\bf 111}, 715 (2020)
[{\it Pis'ma v ZhETF}, {\bf 111}, 833 (2020)].


\bibitem{16} A.G. Lebed, {\it Mod. Phys. Lett. B} {\bf 34}, 2030007
(2020).

\bibitem{Singleton-1}
J. Singleton, J.A. Symington, M.-S. Nam, A. Ardavan, M. Kurmoo,
and P. Day, J. Phys. Condens. Matter, \textbf{12}, L641 (2000).

\bibitem{Tanatar-1}
M.A. Tanatar, T. Ishiguro, H. Tanaka, and H. Kobayashi, Phys. Rev.
B \textbf{66}, 134503 (2002).

\bibitem{Ishiguro-1}
Y. Shimojo, T. Ishiguro, H. Yamoji, G. Saito, J. Phys. Soc. Jpn.
\textbf{71}, 1716 (2002).

\bibitem{Wosnitza-1}
R. Lortz, Y. Wang, A.Demuer, P.H.M. Bottger, B. Bergk, G.
Zwicklnagl, Y. Nakazawa, and J. Wosnitza, Phys. Rev. Lett.
\textbf{99}, 187002 (2007).


\bibitem{Agosta-2}
K. Cho, B.E. Smith, W.A. Coniglio, L.E. Winter, C.C. Agosta, and
J.A. Schlueter, Phys. Rev. B \textbf{79}, 220507 (2009).

\bibitem{Wosnitza-2}
B. Bergk, A. Demuer, I. Sheikin, Y. Wang, J. Wosnitza, Y.
Nakazawa, and R. Lortz, Phys. Rev. B \textbf{83}, 064506 (2011).

\bibitem{Agosta-4}
W.A. Coniglio, L.E. Winter, K. Cho, C.C. Agosta, B. Fravel, and
L.K. Montgomery, Phys. Rev. B \textbf{83}, 224507 (2011).

\bibitem{Brooks-1}
J.A. Wright, E. Green, P. Kuhns, A. Reyes, J. Brooks, J.
Schlueter, R. Kato, H. Yamamoto, M. Kobayashi, and S.E. Brown,
Phys. Rev. Lett. \textbf{107}, 087002 (2011).

\bibitem{Agosta-1}
C.C. Agosta, J. Jin, W.A. Coniglio, B.E. Smith, K. Chao, I. Stroe,
C. Martin, S.W. Tozer, T.P. Murphy, E.C. Palm et al., Phys. Rev. B
\textbf{85}, 214514 (2012).

\bibitem{Kanoda-1}
H. Mayaffre, S. Kramer, M. Horvatic, C. Berthier, K. Miyagawa, K.
Kanoda, and V.F. Mitrovic, Nat. Phys. \textbf{10}, 928 (2014).


\bibitem{Uji-1}
S. Tsuchiya, J.-I. Yamada, K. Sugii, D. Graf, J.S.Brooks, T.
Terashima, and S. Uji, J. Phys. Soc. Jpn. \textbf{84}, 034703
(2015).


\bibitem{Agosta-6}
C.C. Agosta, N.A. Fortune, S.T. Hannahs, S. Gu, L. Liang,
Ju-Hyun Park, and J.A. Schleuter, Phys. Rev. Lett. \textbf{118},
267001 (2017).


\bibitem{Lortz-1}
C.-w. Cho, J.H. Yang, N.F.Q. Yuan, J. Shen, T. Wolf, and R. Lortz,
Phys. Rev, Lett. \textbf{119}, 217002 (2017).

\bibitem{Lortz-2}
C.-w. Cho, J. Lyu, C.Y. Ng, J.J. He, T.A. Abdel-Baset, M. Abdel-Hafiez
and Rolf Lortz, preprint arXiv: 2011.04880 (2020).


\bibitem{Gorkov-1}
A.A. Abrikosov, L.P. Gor'kov, and I.E. Dzyaloshinskii, {\it
Methods of Quantum Field Theory in Statistical Mechanics} (Dover,
New York, 1963).


\bibitem{Gorkov-Lebed}
L.P. Gor'kov and A.G. Lebed, J. Phys. (Paris) Lett. \textbf{45},
L-433 (1984).



\bibitem{Mineev-1}
V.P. Mineev and K.V. Samokhin, {\it Introduction to Unconventional
Superconductivity} (Gordon and Breach Science Publisher,
Australia, 1999).

\bibitem{Gradshein}
I.S. Gradshteyn and I.M. Ryzhik,  {\it Table of Integrals, Series,
and Products} (6-th edition, Academic Press, London, United
Kingdom, 2000).





\bibitem{Gorkov-1}
L.P. Gor'kov, Sov. Phys. JETP, \textbf{37(10)}, 42 (1960).



\bibitem{Lebed-1}
A.G. Lebed, JETP Lett., \textbf{110}, 173 (2019) [Pis'ma v Zh.
Eksp. Teor. Fiz. \textbf{110}, 163 (2019)].



\bibitem{Lebed-2}
A.G. Lebed and O. Sepper, JETP Lett., \textbf{111}, 239 (2020)
[Pis'ma v Zh. Eksp. Teor. Fiz. \textbf{111}, 249 (2020)].


\bibitem{Bula}
L.N. Bulaevskii, Zh. Eksp. Teor. Fiz. \textbf{65}, 1278 (1973)
[Sov. Phys. JETP, \textbf{38}, 634 (1974)].


\bibitem{Bulaevskii}
L.N. Bulaevskii and A.A. Guseinov, Pis'ma Zh. Eksp. Teor. Fiz.
\textbf{19}, 742 (1974) [JETP Lett. \textbf{19}, 382 (1974)].

\bibitem{Klemm}
R.A. Klemm, A. Luther, and M.R. Beasley, Phys. Rev. B \textbf{12},
877 (1975).








\end{references}
\end{document}